\documentstyle[11pt]{article}
\normalbaselines
\begin{document}
TPI-MINN - 98/04

\begin{center}

{\LARGE\bf Quasi-Exactly Solvable Models with Spin-Orbital Interaction}
\footnote{This work was supported by the grant No DOE/DE-FG02-94ER40823}

\vskip 1cm

{\large {\bf Alexander Ushveridze} }

\vskip 0.1 cm
Theoretical Physics Institute,\\
University of Minnesota\\

and\\

Department of Theoretical Physics, University of Lodz,\\
Pomorska 149/153, 90-236 Lodz, Poland\footnote{E-mail
address: alexush@mvii.uni.lodz.pl and alexush@krysia.uni.lodz.pl} \\

\end{center}
\vspace{1 cm}

\begin{abstract}
First examples of quasi-exactly solvable models describing spin-orbital
interaction are constructed. In contrast with other examples of matrix
quasi-exactly solvable models discussed in the literature up to now, our
models admit {\it infinite} (but still incomplete) sets of exact
(algebraic) solutions. The hamiltonians of these models are hermitian
operators of the form $H=-\Delta^2+V_1(r)+({\bf s\cdot l})V_2(r) +
({\bf s+l})\cdot{\bf h}V_3(r)$ where $V_1(r)$, $V_2(r)$ and $V_3(r)$ 
are scalar functions, ${\bf l}$ is a vector of the angular 
momentum operator, ${\bf s}$
is a matrix-valued vector spin-operator and ${\bf {h}}$ is an external
(constant) vector magnetic field. 
\end{abstract}

\section{Introduction}

Quasi-exactly solvable (QES) problems are distinguished by the fact that
only some of their energy levels and corresponding wavefunctions admit
explicit construction. In the last decade a big progress has been acieved in
elaborating the concepts of the phenomenon of quasi-exact solvability and
formulating methods of constructing and solving QES models of a variety of
types (for more detail see e.g. the reviews [\cite{Sh89}, \cite{Us89}, \cite
{GKO94},  \cite{Sh94}] and book [\cite{Us94}]). In this paper we undertake a
new step in this direction and construct new classes of QES models which (up
to now) have never been discussed in the QES-literature. These are QES
models with a spin-orbital interaction. The potentials of such models have
the following general form%
$$
V(x,y,z)=V_1(r)+({\bf s\cdot l})V_2(r)+({\bf j}\cdot {\bf h)}V_3(r) 
$$
where $r=\sqrt{x^2+y^2+z^2}$ is a radial coordinate, ${\bf l}%
=(l_x,l_y,l_z)=(i(y\partial _z-z\partial _y),i(z\partial _x-x\partial
_z),i(x\partial _y-y\partial _x))$ is hermitian (vector) operator of angular
momentum, ${\bf s}=(l_x,l_y,l_z)$ is hermitian (vector) spin-operator whose
components realize a certain unitary finite-dimensional representation of
algebra $so(3)$,  ${\bf j=s+l}$ is the operator of  total momentum,
and ${\bf h}=(h_x,h_y,h_z)$ is an external (constant) magnetic field 
interacting with the total momentum.

One of the most important distinguishing features of QES models which we
intend to present here is that they have an {\it infinite} number of exactly
(algebraically) constructable energy levels and corresponding eigenvalues.
They are however quasi-exactly solvable because the set of their exact
solutions is still incomplete and does not fill all the spectrum of a model.
First examples of such models were presented in our recent work [\cite{LDU}]
where they have been called ''infinite QES models ''.

Another unusual feature of these models is that they (in contrast with
models discussed in paper [\cite{LDU}]) are matrix models with physically
realistic hermitian hamiltonians. Everybody who has some experience with
quasi-exact solvability in the matrix (multi-channel) case knows how
difficult is to satisfy the condition of hermiticity when constructing such
models. It is hardly neccessary to remind the reader that up to now only a
couple of hermitian matrix QES models have been constructed (see e.g. [\cite
{Sh89}, \cite{BrKo94}]).

\section{Starting point}

To demonstrate how does our construction procedure work we start with the
simplest one-dimensional QES model with hamiltonian 
\begin{equation}
\label{a1}H=-\frac{\partial ^2}{\partial r^2}+\frac{(c-1/2)(c-3/2)}{r^2}
+[b^2-2a(2m+c+1)]r^2+2abr^4+a^2r^6
\end{equation}
acting in Hilbert space of functions defined on the positive half axis $r\in
[0,\infty ]$ and vanishing sufficiently fast at its ends $r=0$ and $r=\infty 
$. Here $a,b,c$ are real parameters satisfying the conditions ($a>0,c>0$)
and $m$ is a non-negative integer. As it was demonstrated in [\cite{Us94}],
for any fixed $m$ the Schroedinger equation 
\begin{equation}
\label{a2}H\psi (r)=E\psi (r)
\end{equation}
for model $\ref{a1}$ admits algebraic solutions whose general form is given
by the formulas 
\begin{equation}
\label{a3}\psi (r)=r^{c-1/2}\prod_{i=1}^m\left( r^2/2-\xi _i\right) \exp
\left( -\frac{ar^4}4-\frac{br^2}2\right) 
\end{equation}
\begin{equation}
\label{a4}E=2b(2m+c)+8a\sum_{i=1}^m\xi _i
\end{equation}
The $m$ complex numbers $\xi _i$ in expressions \ref{a3} and \ref{a4}
satisfy the system of $m$ algebraic equations 
\begin{equation}
\label{a5}\sum_{k=1,k\neq i}^m\frac 1{\xi _i-\xi _k}+\frac c{2\xi
_i}-b-2a\xi _i=0,\quad i=1,...,m.
\end{equation}
It turns out  that system \ref{a5} has only $m+1$ permutationally invariant
solutions for any given $m$ which are represented by the sets of real points 
$\xi _i$. Each solution is completely characterized by a (quantum)
number
$k=0,1,...,m$ which indicates the number of positive $\xi _i$-points.
According to formula \ref{a3}, the number of positive $\xi _i$-points
determines the number of (real) wavefunction zeros, which, in turn,
determines the ordinal number of an excitation (oscillation theorem). This
means that model \ref{a1} has $m+1$ exactly constructable solutions
describing the ground state and $m$ first excited states. A more detailed
exposition of properties of model \ref{a1} and its algebraic solutions can
be found in the book [\cite{Us94}].

\section{The modified equation}

It is not difficult to see that the transformation 
\begin{equation}
\label{b1}\psi (r)=r\varphi (r) 
\end{equation}
reduces the equation \ref{a2} to the form 
\begin{equation}
\label{b2}\left( -\frac{\partial ^2}{\partial r^2}-\frac 2r\frac \partial
{\partial r}+\frac{l(l+1)}{r^2}+V(r,l,m)\right) \varphi (r)=E\varphi (r) 
\end{equation}
in which we used the notation 
\begin{equation}
\label{b3}V(r,l,m)=[b^2-2a(2m+5/2+l)]r^2+2abr^4+a^2r^6 
\end{equation}
and 
\begin{equation}
\label{b33}l=c-3/2 
\end{equation}
Hereafter we shall consider $l$ as a new independent parameter taking (by
agreement) only non-negative integer values. The form of the first three
terms in the equation \ref{b2} coincides with the form of the radial part of
a tree-dimensional Laplace operator. For this reason it seems quite natural
to interpret $l$ as the 3-dimensional angular momentum and try to relate the
equation \ref{b2} to a certain 3-dimensional quantum problem. In the
following three sections we show that there are three such possibilities
leading to three different kinds of quasi-exactly solvable problems in the
3-dimensional space.

\section{The first possibility}

One of the simplest possibilities of interpreting equation \ref{b2} is based
on the assumption that function \ref{b3} entering into \ref{b2} is $l$
independent: 
\begin{equation}
\label{c1}V(r,l,m)=V_0(r,N)=[b^2-2a(N+5/2)]r^2+2abr^4+a^2r^6
\end{equation}
For this the number 
\begin{equation}
\label{c2}N=l+2m
\end{equation}
must be fixed. In this case, equation \ref{b2} takes the form of a typical
radial Schroedinger equation for a spherically symmetric 3-dimensional
equation 
\begin{equation}
\label{c3}\left( -\Delta +V_0(r,N)\right) \Psi (x,y,z)=E\Psi (x,y,z).
\end{equation}
Since both $m$ and $l$ are assumed to be positive, the condition \ref{c2}
leads to a finite number of possibilities with $m=0,1,...,[N/2],$ and 
$l=N,N-2,...,N-2[N/2],$ respectively. For this reason, for any given $N,$ the
model \ref{c3} is quasi-exactly solvable and has (as usually) only a finite
( $[N/2]([N/2]+1)/2$) number of explicit solutions. The model of such a form
and even its more complicated spherically non-symmetric versions were
considered many years ago in papers [\cite{Us88b}, \cite{Us89}].

\section{The second possibility}

Another possibility of interpreting equation \ref{b2} is to consider $m$ as
a fixed number not restricting the value of $l$. In this case the function 
\ref{b3} becomes linearly dependent on $l$ and can be represented in the
form 
\begin{equation}
\label{d1}V(r,l,m)=V_1(r,m)-l\cdot
V_2(r)=\{[b^2-2a(2m+5/2)]r^2+2abr^4+a^2r^6\}-l\cdot \{2ar^2\}
\end{equation}
It is quite obvious that, in order to associate the equation \ref{d1} with a
certain 3-dimensional Schroedinger equation, we must find a proper
3-dimensional source for the term which is linear in the momentum $l$. The
first think which comes in ones head is to look for the 3-dimensional scalar
operators $O$ which wuld commute with both the Laplace operator and $r$ and
would have the eigenvalues linear in $l$. In this case we could consider \ref
{b2} as a reduction of a 3-dimensional problem 
\begin{equation}
\label{d2}\left( -\Delta +V_1(r,m)-O\cdot V_2(r)\right) \Psi (x,y,z)=E\Psi
(x,y,z)
\end{equation}
The linearity in $l$ means that the operator must be proportional to the
operator of the angular momentum 
\begin{equation}
\label{d3}{\bf l}=(l_x,l_y,l_z)=(i(y\partial _z-z\partial _y),i(z\partial
_x-x\partial _z),i(x\partial _y-y\partial _x)).
\end{equation}
But this is a 3-dimensional vector while the operator we are looking for
must be a scalar. The only possibility to construct a scalar from \ref{d3}
is to take a scalar product of ${\bf l}$ with another vector operator. It is
quite obvious that there is no such operator if we restrict ourselves to the
single-channel problems. However, if we admit the consideration of
multi-channel problems, then a good candidate for the second operator can
immediately be found. This is obviously the spin operator ${\bf s}$!
Rerstricting ourselves (for the sake of simplicity) to the $1/2$-spin case
(2 by 2 matrices), we can easily check that the spectrum of the operator 
\begin{equation}
\label{d4}O=2\cdot {\bf s\cdot l}
\end{equation}
(which, obviously commutes with both $\Delta $ and $r$) is linear in $l$.
Indeed, representing operator \ref{d4} in the form 
\begin{equation}
\label{d5}O={\bf j}^2-{\bf l}^2-{\bf s}^2
\end{equation}
(where ${\bf j=l+s}$ is a total momentum) and taking for concreteness a
particular case with $j=l+1/2$ we easily find the corresponding branch of
the spectrum 
\begin{equation}
\label{d6}o=j(j+1)-l(l+1)-s(s+1)=(l+1/2)(l+3/2)-l(l+1)-3/4=l.
\end{equation}
This finally leads us to a 3-dimensional matrix QES  models  
\begin{equation}
\label{d9}\left( -\Delta +\{[b^2-2a(2m+5/2)]r^2+2abr^4+a^2r^6\}-
2({\bf s\cdot l})\cdot \{2ar^2\}\right) \Psi (x,y,z)=E\Psi (x,y,z)
\end{equation}
describing spin-orbital interaction.

It is a time to ask ourselves of what kind of models did we obtain? First of
all, one should stress again that these models are actually quasi-exactly
solvable. This follows from the fact that for any given $m$ and $l$ they
have an infinite number of normalizable solutions, but only $m+1$ of them
are exactly (algebraically) constructable. Second, and this is may be the
most important thing, despite the fact that the set of exactly constructable
solutions is incomplete, this set is infinitely large. This is so because
the number $l$ is not fixed by the 3-dimensional model \ref{d9}. It appears
as a solution of the eigenvalue problem for operators $O$ and may take
arbitrary non-negative integer values.

In conclusion of this section note that the hamiltonians of models we
obtained are hermitian by construction.

\section{The third possibility}

The last interesting possibility of reducing the equation \ref{b2} to a
3-dimensional form appears when the function \ref{b3} depends on both
parameters $l$ and $m.$ In this case the function \ref{b3} becomes linearly
dependent on both $l$ and $m$ and can be represented in the form 
\begin{equation}
\label{e1}V(r,l,m)=V_1(r)-l\cdot V_2(r)-m\cdot
V_3(r)=\{(b^2-5a)r^2+2abr^4+a^2r^6\}-l\cdot \{2ar^2\}-m\cdot \{4ar^2\} 
\end{equation}
By analogy with the previous section we can consider the numbers $l$ as the
eigenvalues of the operator of spin-orbital interaction, and the only thing
which remains to do is to interpret $m$ as an independent quantum number
appearing in equation \ref{b2} as an eigenvalue of a certain operator $M$
commuting with the variable $r$, Laplasian $\Delta $ and the spin-orbital
operator ${\bf s\cdot l}$. A good candidate for such an operator is the
z-projection of the total momentum ${\bf s+l.}$ In fact, it should not
neccessarily be a z-projection. Because of the spherical symmetry, it could
be equally weel a x- or y- projection, or any other projection. We can
therefore represent this operator in a covariant form 
\begin{equation}
\label{e2}M=2({\bf s+l})\cdot {\bf h} 
\end{equation}
where ${\bf h}$ is a unit magnetic field. We introduced an additional factor
2 to make the eigenvalues of operator $M$ integer rather than half integer.
Of course, the negative integers are not interesting for us, because, as we
remember, only for non-negative integer values of $m$ the system admits
algebraic solution. Summarizing, we can consider \ref{b2} as a reduction of
a 3-dimensional problem 
\begin{equation}
\label{e3}\left( -\Delta +V_1(r)-\left( {\bf s\cdot l}\right) 
\cdot V_2(r)-2({\bf s+l})\cdot {\bf h\cdot }V_3(r)\right) 
\Psi (x,y,z)=E\Psi (x,y,z) 
\end{equation}
which can be treated as a spectral problem for a matrix quantum model
describing spin-orbital interaction together with the interaction of a total
momentum with an external magnetic field. It is remarkable, that the model 
\ref{e3} does not contain any integer parameters anylonger. All these
parameters appear dynamically as solutions of the eigenvalue problems for
additionally introduced symmetry operators. At the same time, the
model \ref{e3} remains quasi-exactly solvable, because for any 
particular values of
these eigenvalues the equation \ref{b2} has only a certain incomplete set of
solutions.

\section{Conclusion}

The method of construction infinite (matrix) QES models exposed in this
paper is, obviously, quite general and can easily be used for building other
spin-orbital models with more complicated potentials and higher matrix
dimensions. For this it is sufficient to start with other known
one-dimensional QES models first rewritting them in the form of a radial
Schroedinger equation and then interpreting the $l$-dependent terms
appearing in their potential as the eigenvalues of a spin-orbital operators.

\section{Acknowledgements}

I would like to express my sincere gratitude to my colleagues from the
Theoretical Physics Institute of the University of Minnesota (where this
work has been written) for their kind hospitality. I am especially grateful
to Professor M. Shifman for very intersting and fruitful discussions during
my visit. This work was supported by the grant DOE/DE-FG02-94ER40823.


\begin{thebibliography}{9}
\bibitem{LDU}  H. Doebner, K. {\L}azarow  and A. Ushveridze {\it Infinite 
Quasi-Exactly Solvable models}, preprint hep-th No 9707254, 1997


\bibitem{BrKo94}  Y. Brihaye and P. Kosi\'nski, 
{\it Quasi-exactly solvable $ 2\times 2$ matrix equations}, 
J. Math. Phys. {\bf 35}, 3089-3098 (1994)

\bibitem{GKO94}  A. Gonz\'alez-L\'opez , N. Kamran and P.J. Olver, 
{\it Quasi-exact solvability}, Contemp. Math. {\bf 160} 113-140 (1994)

\bibitem{Sh89}  M.A. Shifman, {\it New findings in quantum mechanics:
partial algebraization of the spectral problem}, Int J. Mod. Phys. A 
{\bf 4}, 2897-2952 (1989)

\bibitem{Sh94}  M.A.Shifman, {\it Quasi-exactly solvable spectral problems
and conformal field theory}, Contemporary Mathematics {\bf 160}, 237-262
(1994)

\bibitem{Us88b}  A.G. Ushveridze, {\it Exact solutions of one- and
multi-dimensional Schr\"odinger equations}, Sov. Phys. - Lebedev Inst. Rep. 
{\bf 2} 54-58 (1988)

\bibitem{Us89}  A.G. Ushveridze, {\it Quasi-exactly solvable models in
quantum mechanics}, Sov. J. Part. Nucl. {\bf 20}, 504-528 (1989)

\bibitem{Us94}  A.G. Ushveridze, {\it Quasi-exactly solvable problems in
quantum mechanics}, IOP Publishing: Bristol 1994

\bibitem{Us96}  A.G. Ushveridze, {\it Quasi-exact solvability in local field
theory}, Preprint hep-th No 9607080 (1996)
\end{thebibliography}
\end{document}